\documentclass[aps,pra,reprint,showpacs,twocolumn,longbibliography,nofootinbib]{revtex4-1}%superscriptaddress

\usepackage{graphicx}% Include figure files
\usepackage{dcolumn}% Align table columns on decimal point
\usepackage{bm}% bold math
\usepackage{braket}
\usepackage{leftidx}
\usepackage{cancel}
\usepackage{amsmath}
\usepackage{epstopdf}
\usepackage{color}
\usepackage[mathcal]{eucal}
\usepackage{float}
\usepackage{url}
\usepackage{hyperref}

\usepackage{dcolumn}% Align table columns on decimal point
\usepackage{bm}% bold math
\begin{document}

%\preprint{APS/123-QED}

\title{Quantization of magnetic flux and electron-positron pair creation}% Force line breaks with \\

\author{Mehmet Emre Tasgin} %\thanks{Correspondence: metasgin@hacettepe.edu.tr }
%\email{metasgin@hacettepe.edu.tr}
\affiliation{Institute  of  Nuclear  Sciences, Hacettepe University, 06800 Ankara, Turkey}
\affiliation{metasgin@hacettepe.edu.tr {\rm and} metasgin@gmail.com}

\date{\today}% It is always \today, today,

\begin{abstract}
An electron-positron~($e^- \: e^+$) pair is created in vacuum above a critical electric field strength $E_{\rm crt}$ which is quite large in the laboratory scale. The photon, thus the field, annihilates in the pair creation process. Here, we question if the pair creation (at $E=E_{\rm crt}$) introduces a boundary condition in the electromagnetic state, e.g., similar to the one in angular or linear momentum. We show that introduction of such a reasonable condition yields approximately the well-known magnetic flux quanta which normally one obtains using different arguments. 
\end{abstract}

\maketitle

%\tableofcontents

%\section{\label{sec:level1}Introduction}

%\section{Introduction}

In 1951, Swinger showed that a strong static electric field can create an electron-positron~($e^- e^+$) pair from the quantum electrodynamics~(QED) vacuum~\cite{schwinger1951gauge,reuter1997constant}. The pair creation probability $\propto \sum_{n=1}^\infty  1/n^2 \exp(-n\pi E_{\rm swg}/E )$ for the lightest charged particle ---electron--- has an exponential dependence on the electric field strength $E_{\rm swg}=m_e c^3/e\hbar\simeq 1.3\times 10^{18}$ V/m, where $m_e$ is the electron mass~\cite{gelis2016schwinger}. The $e^- e^+$ pair production probability is appreciable for electric field strengths $E\sim E_{\rm swg}$.

Developments in the X-ray free electron laser~(XFEL) technology pawed the way to the generation of electric field strengths as large as $E_{\rm \scriptscriptstyle XFEL}\sim 0.1\times E_{\rm swg}$ in the laboratories~\cite{ringwald2001pair}. This made the direct production of the $e^- e^+$ pairs from vacuum a hot topic in the past two decades~\cite{alkofer2001pair,roberts2002quantum,narozhny2004PLA,oluk2014electron,li2017momentum,olugh2019pair,ruf2009pair,kohlfurst2020effect,kohlfurst2016effect,bulanov2010multiple,
fried2001pair,kohlfurst2014effective,schmidt1998quantum}. Further (improved) theoretical studies~(e.g., Refs.~\cite{fried2001pair,bulanov2010multiple}) showed that field strengths in dimensions of $E_{\rm crt}\sim 0.1\times E_{\rm swg}$ can also achieve appreciable $e^- e^+$ pair creation~\cite{alkofer2001pair,roberts2002quantum,narozhny2004PLA,oluk2014electron,ringwald2001pair,olugh2019pair,ruf2009pair}. Pair creation, taking place at the critical field strength, annihilates the photon thus wipes out the electric field. 

A second phenomenon, which looks not connected with the Schwinger pair creation process at the present, is the quantization of the magnetic flux in charged superfluids, i.e., superconductors~(SCs)~\cite{Kittel2004}. The flux quantization appears, because the wavefunction $\psi({\bf r}) = n^{1/2} e^{i\theta({\bf r})}$ governing the superfluid Cooper pairs must be single valued at each contour, i.e., $\psi(\theta=0)=\psi(\theta=2\pi)$. A wavefunction of the same form, $\psi({\bf r}) = n^{1/2} e^{i\theta({\bf r})}$, appears also for other superfluid media~\cite{ginzburg2009theory}, e.g., Bose-Einstein condensates~\cite{pethick2008bose}, where quantization phenomena take place due to a similar condition~\footnote{The collective behavior of constituent particles~\cite{tasgin2017many} in these systems enables a single wavefunction description of the system.}.

A quantization phenomenon appears between canonically conjugate operators. Quantization of angular momentum, $\hat{L}_z \doteq \frac{\hbar}{i}\frac{\partial}{\partial \phi}$, is a constraint which has to be satisfied by an arbitrary system. It occurs due to the single valuedness of the wavefunction under azimuthal rotations, i.e.,  $\psi(\phi$=$0$)$=\psi(\phi=2\pi)$. A similar condition, $\psi(x=0)=\psi(x=L)$, quantizes the linear momentum~($\hat{p}\doteq \frac{\hbar}{i} \frac{\partial}{\partial x}$) in an optical cavity,  condensed matter physics~\cite{Kittel2004}, or optical lattices~\cite{morsch2006dynamics}. (Here, $L=Na$ in a lattice, with $a$ is the spatial periodicity and $N$ is the number of sites.) We note that, the operators $\hat{x}$ and $\hat{p}$, or $\hat{\phi}$ and $\hat{L}_z$, are canonically conjugate to each other~\cite{SakuraiBook}.

Similarly, electric $\hat{E}$ and magnetic $\hat{H}$ field operators are also canonically conjugate to each other for a given electromagnetic mode~\cite{mandel_wolf_1995}. In quantum optics, one can represent the electromagnetic~(EM) quantum state in the $E$ field basis ---named as the $q$-representation~\cite{mandel_wolf_1995}--- as well as using the the Fock or coherent state bases. For instance, in the vacuum state the probability for a particular field strength $E$ to occur is $|\psi(E)|^2 \propto \exp(-E^2/\varepsilon_k^2)$ for a mode of frequency $\omega_k$. Here, $\varepsilon_k=\left(   \frac{\hbar \omega_k}{2V\epsilon_0}\right)^{1/2}$ and $V$ is the quantization volume. Thus, in analogy with $\hat{p}$ and $\hat{L}_z$, one can represent the magnetic field operator as $\hat{H_y}\doteq \frac{\hbar}{i} \frac{\omega_k c}{V} \frac{\partial}{\partial E_x}$ for a given mode.

In this paper, we raise the following question. Is it possible that magnetic flux quantization phenomenon ---normally obtained using totally different arguments--- is somehow connected with the take place of $e^- e^+$ pair creation above a critical electric field strength? More explicitly, does annihilation of the photon (so the electric field) above a critical field strength $E= E_{\rm crt}$ introduce the EM wavefunction an analogous boundary condition, e.g., like $\psi(0)=\psi(E_{\rm crt})$ ?  In this work, we obtain a magnetic flux quantum $\Phi_{E_{\rm crt}}$ using the pair creation as a boundary condition. The flux quantum we obtain from the pair creation condition $\Phi_{E_{\rm crt}}\sim 2.3\times 10^{-15}$ T$\times {\rm m}^2$ is quite close to the well-known magnetic flux quanta $\Phi_0\simeq 2.0978\times 10^{-15}$ T$\times{\rm m}^2$ obtained in a superconducting loop using another condition $\psi(\theta=0)=\psi(\theta=2\pi)$~\cite{Kittel2004}.

{\bf \small Flux quantization due to pair-creation}.--- We use plain arguments for obtaining the $\Phi_{E_{\rm crt}}$. Rephrasing ourselves, an EM quantum state creates an $e^-e^+$ pair in the QED vacuum above the critical field strength $E_{\rm crt} \sim 0.1\times E_{\rm swg}$~\cite{alkofer2001pair,roberts2002quantum,narozhny2004PLA,oluk2014electron,li2017momentum,bulanov2010multiple,fried2001pair}. The photon, thus the field, is annihilated when the pair is created. This phenomenon is possible to introduce a boundary condition on the electric field. Thus, a quantity including the $\hat{H}_y$ operator, which is canonically conjugate to $\hat{E}_x$, is possible to be quantized. The first guess, naturally, is the magnetic flux. So, we calculate the quantity
\begin{equation}
\Phi_{E_{\rm crt}} = \mu_0 H_y A = n\times \mu_0 \frac{\omega_k c}{V} \frac{2\pi}{E_{\rm crt}} \hbar A
\end{equation}
within the pair creation region. Here,  $n=0,1,2...$ is an integer and $V$ is the quantization volume in which the pair production takes place. $A$ is the approximate area the $H_y$ field faces within the pair creation region. In QED, the complementary~(its Hermitian conjugate, or the reverse) process takes place simultaneously: electron-positron annihilation creates a photon in the region measured with a Compton wavelength~\cite{fried2001pair,StricklandBook2019,erber1966high,saakyan1960single}~\footnote{One should note that here we do not study the pair production from an XFEL where quantization volume is very different. Here, we consider an annihilation or (reverse) creation process which takes place in a region of Compton wavelength dimensions.}.

The flux quantization due to the pair creation can be re-expressed as
\begin{equation}
\Phi_{E_{\rm crt}} = n \times \mu_0 \frac{\hbar\omega_k c}{L} \frac{2\pi}{E_{\rm crt}}.
\end{equation}
We consider a photon frequency ---for the one annihilated in the pair creating process--- which matches the pair energy  $\hbar\omega_k \sim 2 m_e c^2$. We also use the Compton wavelength of the annihilated photon (or the generated one in the reverse process) for the length $L\sim \lambda_{\rm comp}=h/2m_ec$. Hence, for $E_{\rm crt}\sim 0.1\times E_{\rm swg}$, we obtain a magnetic flux quantum of
\begin{equation}
\Phi_{E_{\rm crt}} = n\times 2.3 \times 10^{-15} \quad {\rm T}\times{\rm m}^2
\end{equation}
using the pair creation condition  on the electric field. This is quite close to the magnetic flux quanta
\begin{equation}
\Phi_0=h/2e=2.068\times 10^{-15} \quad {\rm T}\times{\rm m}^2
\end{equation}
one obtains from the condition $\psi(\theta=0)=\psi(\theta=2\pi)$ in a closed loop of a charged superfluid~\cite{liberati2014astrophysical}.

{\bf \small Brief discussions}.---
We also need to mention that studies on $e^-e^+$ pair creation phenomenon show deviations between different models or depending on the approximations used~\cite{bulanov2010multiple,fried2001pair,ruf2009pair,kohlfurst2020effect,kohlfurst2016effect,kluger1998quantum,rau1996reversible}.
For instance, some of the works~(e.g., Refs.~\cite{li2017momentum,olugh2019pair}) employ the field $E_{\rm crt}\sim 0.1 \sqrt{2} E_{\rm swg}$ in their studies. Using this value as the boundary condition, we obtain a magnetic flux quantum $\Phi_{E_{\rm crt}} = n\times 1.7 \times 10^{-15} \quad {\rm T}\times{\rm m}^2$ which is slightly below the $\Phi_0$. We further note that our calculation regarding the flux quantum also includes imprecise estimates which actually necessitates careful QED calculations. For instance, we calculate the magnetic flux simply using the area $A$, which we refer as the one in which pair creation/annihilation takes place. It is similar for using $L\sim \lambda_{\rm comp}$. 

While the match between the two calculations employing such ``large'' and ``small'' numbers is impressive, it can also be a coincidence. Nevertheless, taking into account that the match between the two fluxes relies on a reasonable condition ---$e^-e^+$ pair creation above $E_{\rm crt}$--- it deserves further investigations. We emphasize that even a 1-order of magnitude match between the two phenomena would be  engrossing enough for triggering further studies.

{\bf \small Consequences}.--- We discuss that the well-known magnetic flux quantization is possible to occur also due to the $e^- e^+$ pair creation process. This phenomenon may refashion our understanding on the nature of the space, fields and charged particles. For instance, such an observation may relate the azimuthal feature of the universe, i.e., $\psi(\theta=0)=\psi(\theta=2\pi)$, the elementary charge $q=2e$, and the pair creation phenomenon (above $E_{\rm crt}$) to each other in a further research.

If the phenomenon we present is not a coincidence, the result we obtain makes one of the universal parameters, like mass or charge of an electron, depend on the other ones. This is like, what happens after one realizes the $\mu_0 \epsilon_0=1/c^2$ relation. One can express magnetic permeability as $\mu_0=1/c^2 \epsilon_0$.

%\begin{figure}
%\centering
%\includegraphics[width=0.5\textwidth]{fig2} 
%\caption{ }
%\label{fig2}
%\end{figure}

%\begin{acknowledgments}
{\bf \small Acknowledgments}.---
We gratefully thank Vural G\"{o}kmen for his motivational support and Bayram Tekin and Ceyhun Bulutay for the scientific support. 
%\end{acknowledgments}

\bibliography{bibliography}% Produces the bibliography via BibTeX.

\end{document}